\documentclass[9pt,twocolumn,twoside]{osajnl}
\journal{ol} 

\setboolean{shortarticle}{true}

\title{Molecular cooling via Raman transitions and a harmonic spectral modulation of an optical frequency comb: The role of the parity of  the chirp in quantum control}

\author[*]{S. A. Malinovskaya}
\author{G. Liu}

\affil{Department of Physics and Engineering Physics, Stevens Institute of
Technology, Hoboken, NJ 07030}

\affil[*]{Corresponding author: smalinov@stevens.edu}

\dates{Compiled \today}

\ociscodes{(320.7110 ) Ultrafast nonlinear optics; (270.0270) Quantum optics.}

\doi{\url{http://dx.doi.org/10.1364/ao.XX.XXXXXX}}

\begin{abstract}
A method for creation of ultracold molecules by stepwise adiabatic passage from the Feshbach state to the fundamentally ground state using an optical frequency comb is presented within a semiclassical multilevel model. The sinusoidal modulation of the  spectral phase of the comb is implemented that leads to a creation of a quasi-dark dressed state having an insignificant population of the excited state manifold and, thus, efficiently mitigating decoherence in the system. In contrast, the cosine modulation does not lead to the quasi-dark state formation. The results demonstrate the importance of the parity of the spectral chirp in quantum control. 
\end{abstract}

\setboolean{displaycopyright}{true}

\begin{document}

\maketitle
\thispagestyle{fancy}

\ifthenelse{\boolean{shortarticle}}{\abscontent}{}
\section{Introduction}

Ultracold control has originated on the base of latest developments in the field of ultracold gases, that
opened up possibilities to generate ultracold molecular systems that offer internal structure and exhibit long-range, dipole-dipole interactions \cite{La09}. The development of methods for molecular cooling continues to be an actual objective owing to diversified internal structure in molecules limiting the implementation of a well established stimulated Raman adiabatic passage (STIRAP) \cite{Ni08}. Quantum control methods have demonstrated high potential to produce ultracold molecules and open up a wide range of opportunities to operate on a time scale much faster than the spontaneous decay \cite{Metcalf}. 

 Decoherence is inherently present in ultracold dynamics and reveals itself in quantum measurements. A study of decoherence is of particular importance for the development of  methods to manipulate ultracold atomic gases, create and control ultracold molecules, for the advancements in ultracold chemistry, quantum computation science, etc.  At ultracold temperatures, decoherence acquires new features while occurs in matter systems free from thermal motion. Under these conditions, the dephasing due to collisions is not longer a limiting factor, \cite{Cu05}, in contrast to this effect at room temperatures \cite{Eberly}. 
%
%
The central idea to success of the control
scenarios in systems with decoherence is maintenance of the phase information contained both in
matter and light. Understanding of the impact of decoherence, thus, is crucial for experimental realizations. %




In this letter, we propose a quantum control method for a preparation of molecules in an ultracold state, e.g., the ground electronic, rovibrational, spin-singlet state, starting from the Feshbach state by means of two-photon Raman transitions induced by a spectrally modulated ultrafast pulse train. We will analyze decoherence effects that lead to
the loss of the quantum phase and have detrimental effects on controllability.

\section{Theoretical framework}
\label{sec:examples}
\begin{figure}
\centerline{\includegraphics[width=5cm]{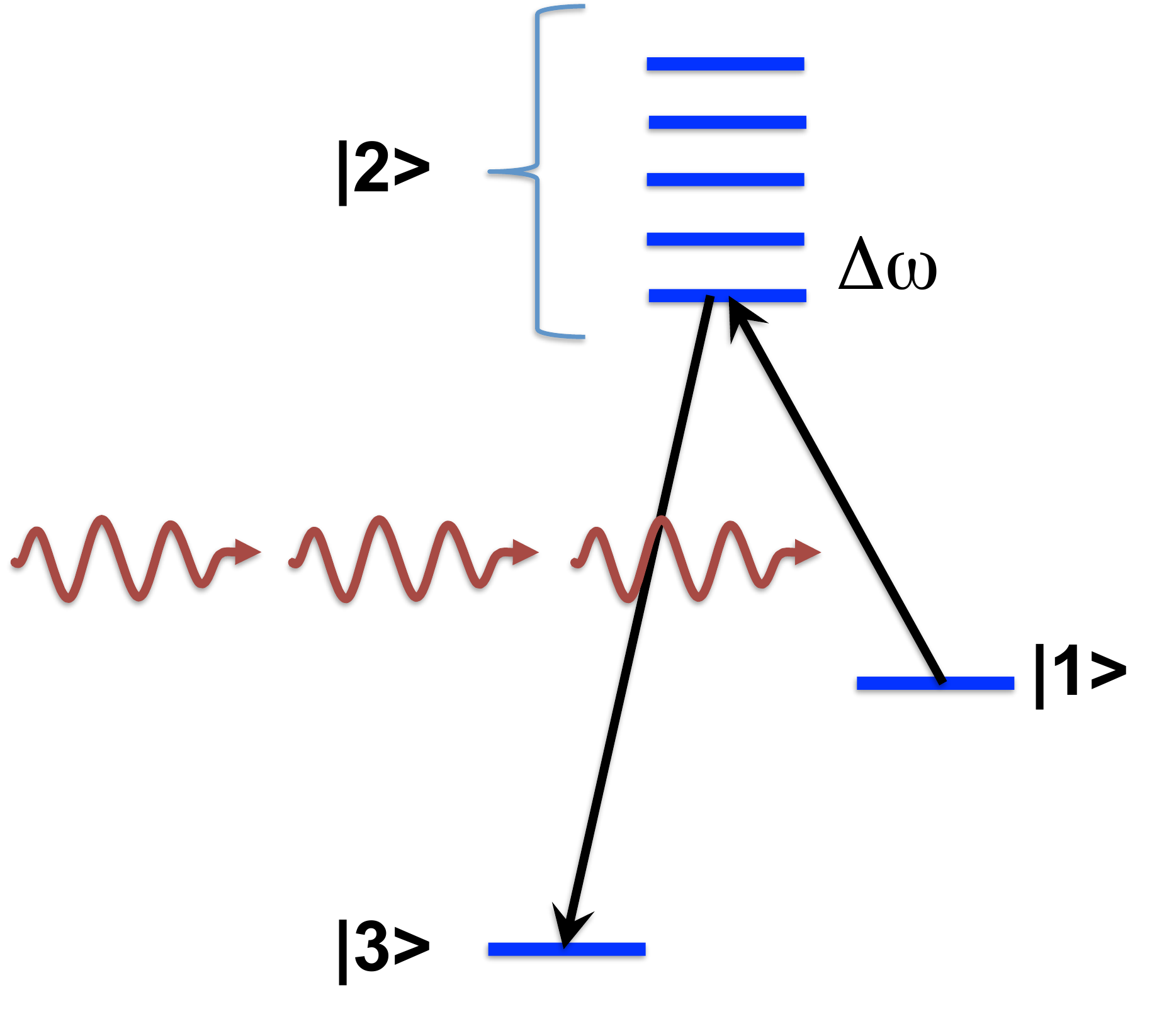}}
\caption{A multi-level $\Lambda$ system modeling the Feshbach state $|1>$, the excited state manifold, $|2>$  and the final state $|3>$ in a molecule interacting with a phase locked pulse train. The transitional frequencies used in calculations are $\omega_{21}$ = 340.7 THz, $\omega_{32}$ = 410.7 THz, and $\omega_{31}$=70. THz  \cite{Sh08}, or $\omega_{21}$=309.3 THz,  $\omega_{32}$=434.8 THz, and $\omega_{31}$=125.5 THz \cite{Ni08}. The $\Delta \omega$ = 0.1$\omega_{31}$  }\label{seven_scheme_FC}
\end{figure}

We consider a semiclassical model of a seven-level quantum system interacting with a classical phase-locked pulse train. Each pulse in the train is spectrally modulated in a form of a sinusoidal function. The model aims to describe a controllable internal dynamics in molecules initiated from the Feshbach state and resulting in molecular stepwise transition into the ultracold state. The seven-level system has a $\Lambda$ configuration formed by the initial Feshbach state, five transitional, electronically excited rovibrational states and the final ultracold state, Fig(\ref{seven_scheme_FC}). The electronically excited rovibrational states are equally separated in accordance with the one-dimensional harmonic oscillator solution, which is a good approximation for the low laying internal states. The frequency separation $\Delta \omega$ between these states corresponds to the vibrational frequency of the totally symmetrical mode in the KRb molecule.  

The sinusoidal modulation of the spectral phase of the incident field is introduced in the frequency domain as $M(\omega)=e^{-i A \sin(\omega T_0 + \phi)}$ \cite{Baumert}. 
 The spectral profile of the field is
\begin{equation}
\begin{aligned}
E(\omega)&=&\sqrt{\pi/2} E_0 \tau (e^{-\frac{1}{2}(\omega-\omega_0)^2\tau^2} + e^{-\frac{1}{2}(\omega+\omega_0)^2\tau^2}) \\
& & \cdot e^{-i A \sin(\omega T_0 + \phi)} \sum \limits_{k=0}^{N} e^{i \omega k T} \label{field_spectr}
\end{aligned}
\end{equation}
Here $E_0$ is the peak field amplitude, $\tau$ is the pulse duration, $T$ is the pulse train period, $\omega_0$ is the carrier frequency, $A$ is the modulation amplitude, $T_0^{-1}$ is the modulation frequency, $\phi$ is the phase, which may be chosen accordingly to address the parity of the chirp. 

The Fourier transform of the modulation function gives the temporal variation of the phase of the field in the form of a series of Bessel functions $M(t)=\sum\limits_{n=-\infty}^{\infty}  J_n(A) e^{-in\phi} \delta(t-nT_0)$. 
The pulse train with such phase modulation reads
\begin{equation}\label{OFC}
\begin{aligned}
E(t) & = & E_{0}\sum\limits _{n=-\infty}^{\infty}J_{n}(A)e^{-in\phi}\sum\limits _{k=0}^{N}e^{-(t-nT_{0}-kT)^{2}/(2\tau^{2})}\\
&  & \cdot\cos{(\omega_{0}(t-nT_{0}-kT))}.
\end{aligned}
\end{equation}

For the description of the time evolution of the seven-level system we refer to the Leouville-von Neumann equation $i\hbar\dot{\rho}=[H_{int},\rho]$  
with relaxation terms that take into account spontaneous decay and elastic collisions. 

Five excited vibrational states are uncoupled since radiationless transitions between vibrational states within a single electronic state are not included. Thus, the system may be treated as a superposition of five three-level $\Lambda$ systems with the excited states differing by $(q-1) \Delta \omega$, where $q$ is an integer from 1 to 5. The interaction Hamiltonian reads $\hat{H}_{int}=  \sum \limits_{q=1}^{5} \hat{H}^q_{int}$, with $ \hat{H}^q_{int} $ describing a single three-level $\Lambda$ system from a set of five of them. The $\hat{H}^q_{int} $ contains terms proportional to the spectral sinusoidal modulation of the field  and, in the rotating wave approximation, has 
two nonzero matrix elements $\hat{H}^q_{ij}=\Omega_R(t) [\exp\{ i
((\omega_0 - \omega_{ji}) - (q-1) \Delta \omega) t),$ here, i,j are the indexes of the
basis set, $i=1,2$ and $j=i+1$,  and the Rabi frequency $\Omega_R(t)= -\mu/\hbar \sum \limits_n J_n(A)E_0 e^{-(t-nT_0)^2/(2\tau_0^2)-i n \omega_0 T_0}$, with the peak value $\Omega_R$. The Leouville-von Neumann equation gives a
set of coupled differential equations for the density matrix elements 

\begin{eqnarray} \label{LvN}
&\dot{\rho}_{11}=2Im[H_{12}\rho_{21}] \nonumber \\
&\dot{\rho}_{22}=2Im[H_{21}\rho_{12}+H_{23}\rho_{32}] \nonumber \\
&\dot{\rho}_{33}=2Im[H_{32}\rho_{23}] \nonumber \\
&\dot{\rho}_{12}=-i H_{12}(\rho_{22}-\rho_{11})+iH_{32}\rho_{13}  \\
&\dot{\rho}_{13}=-iH_{12}\rho_{23}+iH_{23}\rho_{12} \nonumber \\
&\dot{\rho}_{23}=-i H_{23}(\rho_{33}-\rho_{22})-iH_{21}\rho_{13}
 \nonumber
\end{eqnarray}

The decoherence terms are taken into account through the reduced density matrix elements
\begin{equation}
\begin{array}{ccc}
\dot{\rho}_{11})_{sp}=\gamma_{1}\rho_{22}  &
 \dot{\rho}_{12})_{sp,col}=-(
 \frac{\gamma_{1}}{2}+\frac{\gamma_{2}}{2}+\Gamma_{1} )
\rho_{12} & \\
\dot{\rho}_{22})_{sp}=-\gamma_{1}\rho_{22}-\gamma_{2}\rho_{22}&
\dot{\rho}_{13})_{sp,col}=-\Gamma_{3}
\rho_{13} & \\
 \dot{\rho}_{33})_{sp}=\gamma_{2}\rho_{22}  &
\dot{\rho}_{23})_{sp,col}=-(
  \frac{\gamma_{1}}{2}+\frac{\gamma_{2}}{2}+\Gamma_{2} )
\rho_{23}. & \\
\end{array} \label{REDST_DENSITY}
\end{equation}
The decay from the excited states to the initial and the final  state is at the rate $\gamma_1$ and $\gamma_2$, respectively, and that from the initial to the final state is at the rate $\gamma_3$. Collisions are considered between the molecules in any state within the model and the buffer gas, they are described by three dephasing rates, $\Gamma_1$, $\Gamma_2$ and $\Gamma_3$ related to dephasing between the excited state manifold and the initial state, that and the final state and the initial state and the final state, respectively. Dephasing due to collisions between molecules in different vibrational states of the electronically edited states is not taken into account. 
 The reduced density matrix elements in Eqs.(\ref{REDST_DENSITY}) were added to Eqs.(\ref{LvN}), which were solved numerically.

\begin{figure}
 \centerline{
\includegraphics[width=8cm]{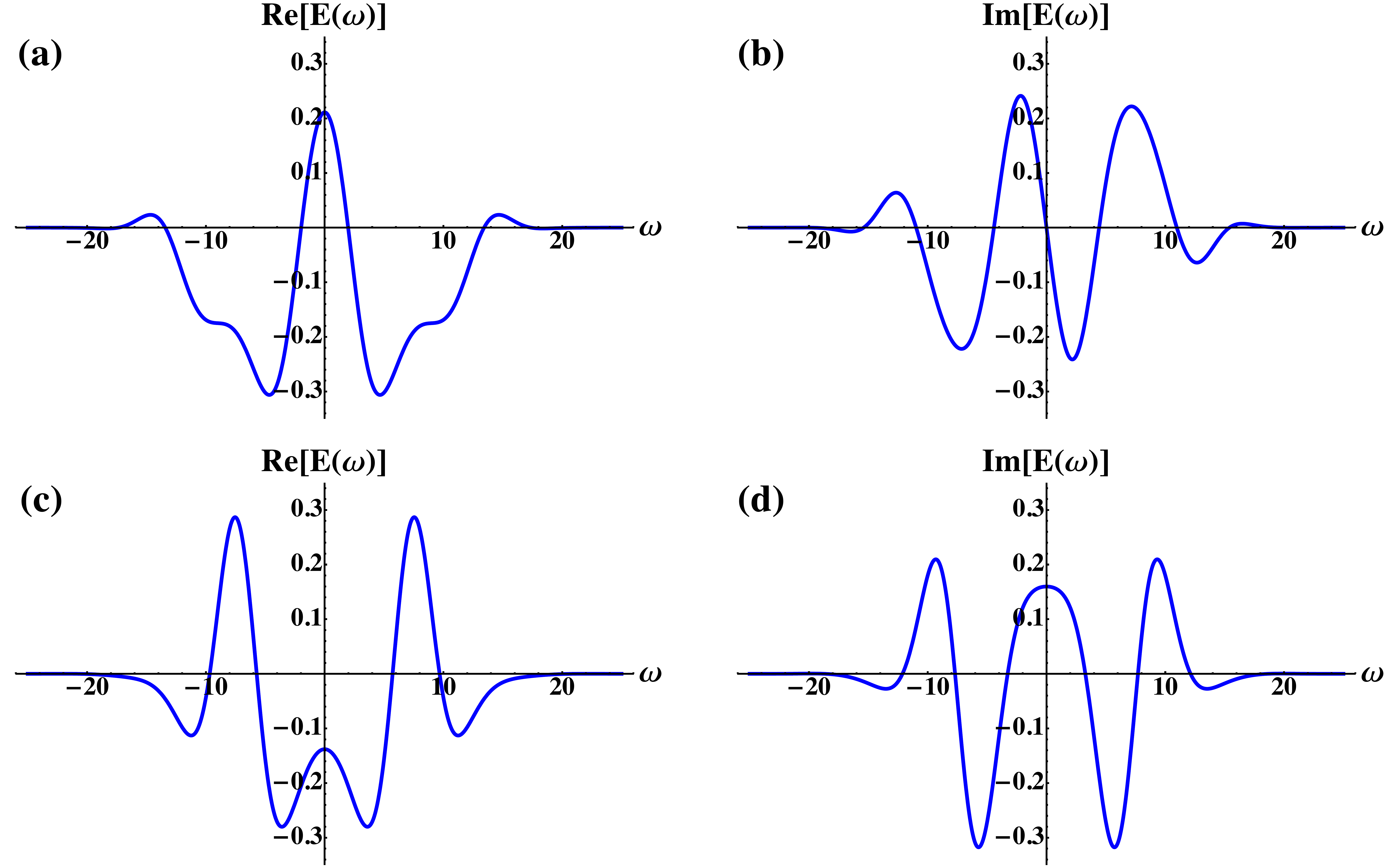}}
\caption{The real and imaginary components of the field $E(\omega)$, Eq (\ref{field_spectr}), for $\phi = 0$ (a,b) and $\phi = \pi/2$ (c,d).}\label{Re_Im_field}
\end{figure}

In the expression for the external field Eq.(\ref{OFC}), the $\phi$ governs the parity of the chirp. 
The real and imaginary components of the field $E(\omega)$, Eq (\ref{field_spectr}), are shown in Fig.(\ref{Re_Im_field}) for the sine and cosine modulation and manifest different phase relationship between different components of the field. For the sine modulation (odd chirp), the imaginary part is asymmetric with respect to zero frequency 
while the real part is symmetric and has components dominantly negative in amplitude. In the case of the cosine modulation (even chirp), both the real and imaginary parts of the spectral field amplitude are symmetric. In any case, a pair of frequencies from the positive and the negative region of the spectrum provides the resultant Raman field having the amplitude that depends on the amplitude difference of two interfering field components. Odd chirp causes cancellation of the field amplitudes being $\pi-$shifted with respect to each other, meantime, the even chirp provides the addition of the imaginary components.  

\section{Numerical results and discussions}

\begin{figure}
 \centerline{
\includegraphics[width=9cm]{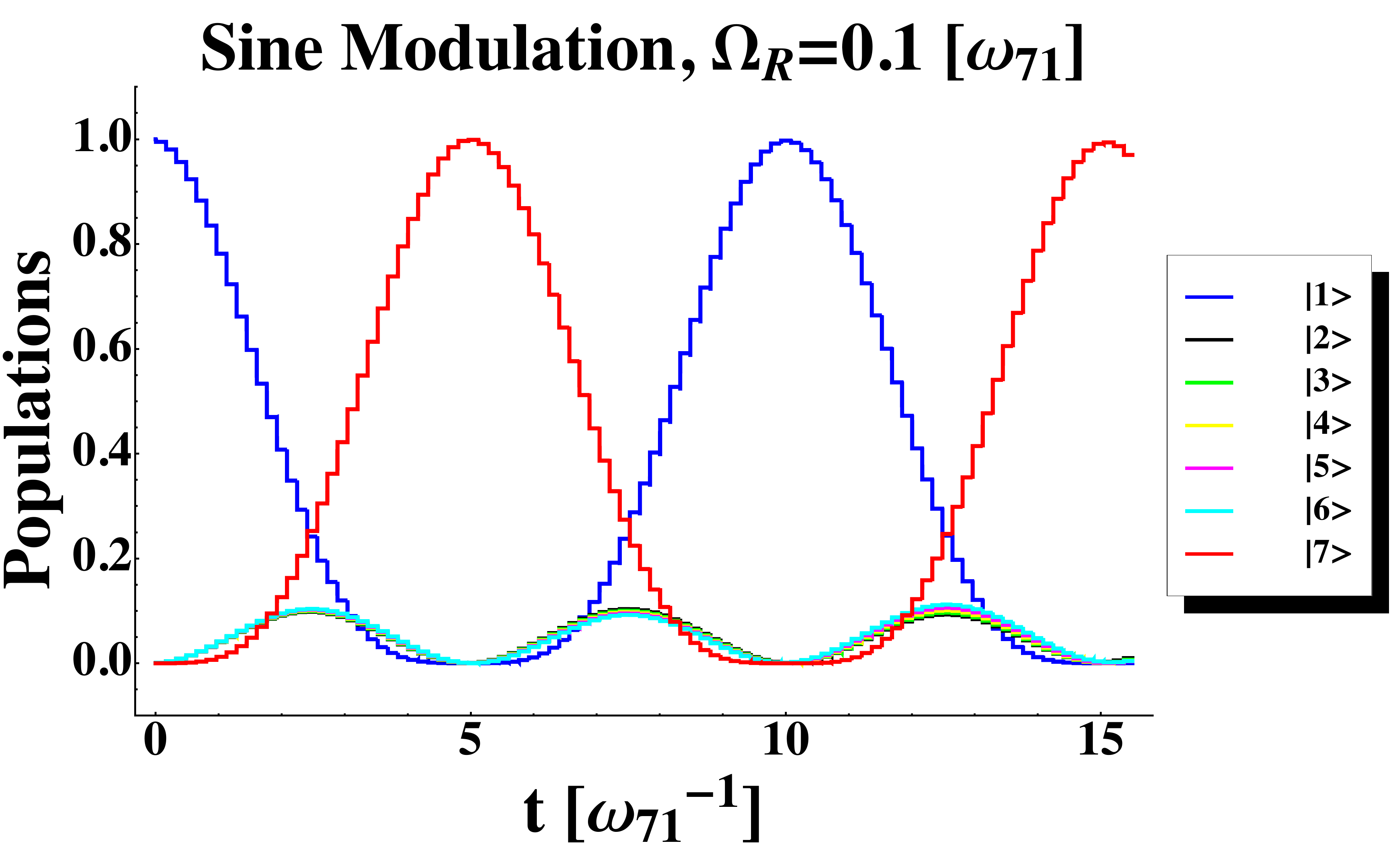}}
\caption{Population dynamics in the seven-level $\Lambda$ system induced by the resonant Raman transitions using a sine-modulated pulse train described by Eq.(\ref{OFC}), ($\phi=0.$). The values of the system parameters are  $\omega_{32}$ = 410.7 THz, $\omega_{21}$ = 340.7 THz, $\omega_{31}$ =70 THz and $\Delta \omega = 0.1\omega_{31}$ \cite{Sh08}. The carrier frequency $\omega_0$ = $\omega_{32}$, the spectral modulation parameter $T_0$ = 1/$\omega_{21}$, the modulation amplitude $A=4$ and the peak Rabi frequency $\Omega_R$=7 THz. A single pulse duration $\tau$ = 3 fs, pulse train period is T = 19.2 ns. Rabi oscillations are observed between the initial Feshbach state and the final ultracold state.
}\label{7lvl_sin}
\end{figure}

Spectrally sine modulated pulse train, when interacts with the seven-level system, Fig.(\ref{7lvl_sin}), produces a peace-wise adiabatic passage of population from the initial Feshbach state and the final ultracold state  
by a series of coherent pulses. The excited state manifold is insignificantly populated transitionally. For each incident pulse, except for the first one, the "initial condition" of the seven-level system is the one, prepared by the previous pulse. It implies nonzero  population of the Feshbach state and all other states. Each pulse in the pulse train transferrs a fraction of population from the initial to the final state. For the peak Rabi frequency same as the two-photon transitional frequency $\omega_{31}$, a complete adiabatic population transfer occurs in 32 pulses. The system parameters are $\omega_{21}$ = 340.7 THz, $\omega_{32}$ = 410.7 THz, and $\omega_{31}$=70. THz  \cite{Sh08}, the $\Delta \omega$ = 0.1$\omega_{31}$. The carrier frequency $\omega_0$ is in the one-photon resonance with the $\omega_{32}$ transitional frequency and the reciprocal of spectral modulation parameter $T^{-1}$ is equal to the transitional frequency $\omega_{21}$. The $\omega_0 - T^{-1}$ satisfies the two-photon resonance condition and equals to $\omega_{31}$. Two complementary mechanisms of the resonant, two-photon adiabatic passage are provided: by the frequency difference $\omega_0 - T^{-1}$, from one hand, and by the difference of the pairs of optical frequencies in the comb, that are multiples of radio frequencies, from another hand. Detuning  of $\omega_0$ and $T^{-1}$ off the one-photon resonances reduces the impact of the first mechanism and results in greater population of the excited state manifold. The second mechanism strongly depends on the period of the pulse train, since it determines the one-photon detuning of the optical frequencies, \cite{Sh10,Ma13}. 
 When $\omega_0 - T^{-1} = \omega_{31}$, dynamics occurs within a quasi-dark state with the excited state manifold insignificantly populated. 
The system response resembles the one in the STIRAP control scheme \cite{Klaas}. 
To gain insight into dynamics of a stepwise population transfer to the final state within the dark state by the spectrally sine modulated pulse train, we performed the dressed state analysis. We approximated the excited state manifold by a single vibrational state of the excited electronic state for a clear picture of the passage. This is a feasible action because the excited states are all similarly populated according to the exact solution of the Schr\"odingier equation for the seven-level system. The Hamiltonian of a single pulse interaction with the three-level system in the field interaction representation reads
\begin{equation} 
\hat{H}_d=\left[\begin{array}{ccc}
0  &
R(t) e^{i (\omega_{31}t +M(t))}& 0\\
R(t) e^{-i (\omega_{31}t +M(t))}&
0 & R^\prime(t) e^{i M^\prime(t)}\\
0  &
R^\prime(t) e^{-i M^\prime(t)} & 0 
\end{array}\right]. \label{HAM_DRESS}
\end{equation}
Here $R(t)e^{iM(t)} = -\frac{\mu}{\hbar}E_0 \sum\limits_{n=-\infty}^{\infty} J_n(A) e^{-in\phi} e^{-\frac{(t-nT_0)^2}{2\tau^2}} e^{-i n\omega_0 T_0}$ and $ R^\prime(t)e^{iM^\prime(t)} = -\frac{\mu}{\hbar}E_0 \sum\limits_{n=-\infty}^{\infty} J_n(A) e^{-in\phi} e^{-\frac{(t-nT_0)^2}{2\tau^2}} e^{i n\omega_0 T_0} $
The time dependence of the dressed state energies is obtained numerically for the carrier frequency $\omega_0$ = $\omega_{32}$, the spectral modulation parameter $T_0$ = 1/$\omega_{21}$, the modulation amplitude $A=4$ and the peak Rabi frequency equal to the transition frequency between the Feshbach and the ultracold state. A single pulse duration is $\tau$ = 3 fs and the three-level pulse train period is T = 19.2 ns. The three-level system parameters are $\omega_{21}$=309.3 THz,  $\omega_{32}$=434.8 THz, and $\omega_{31}$=125.5 THz \cite{Ni08}.

\begin{figure}
\centerline{
\includegraphics[width=9cm]{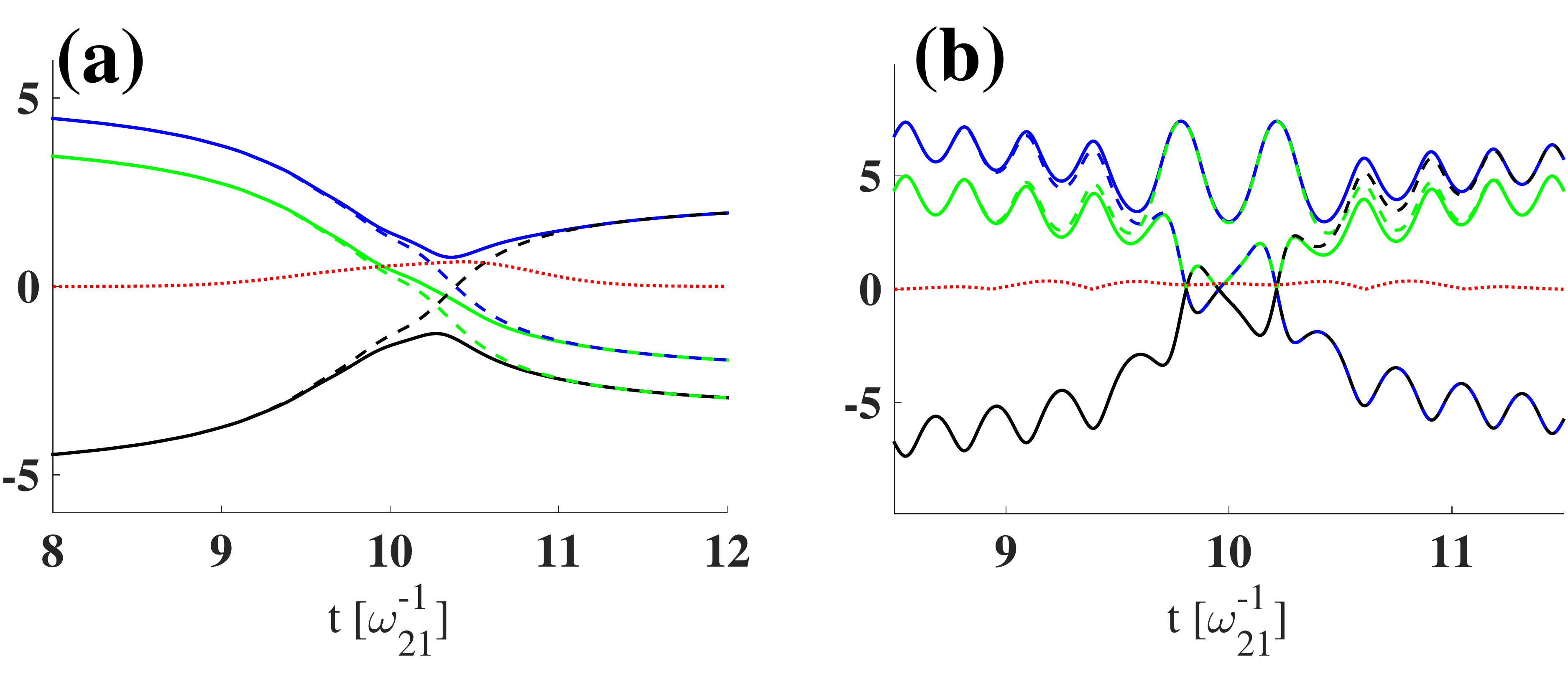}}
\caption{Dynamics of the dressed state energies in the three-level $\Lambda$ system induced by the resonant Raman transitions using a spectrally modulated by a sin (a) and cos (b) function pulse train described by Eq.(\ref{OFC}). 
A single pulse duration $\tau$ in the pulse train is 3 fs, the pulse train period is 19.2 ns. The peak Rabi frequency is equal to the transition frequency between the Feshbach and the ultracold state, $\Omega_R$=125.5 THz. }\label{dress_sinsptr}
\end{figure}

\begin{figure}
\centerline{
\includegraphics[width=9cm]{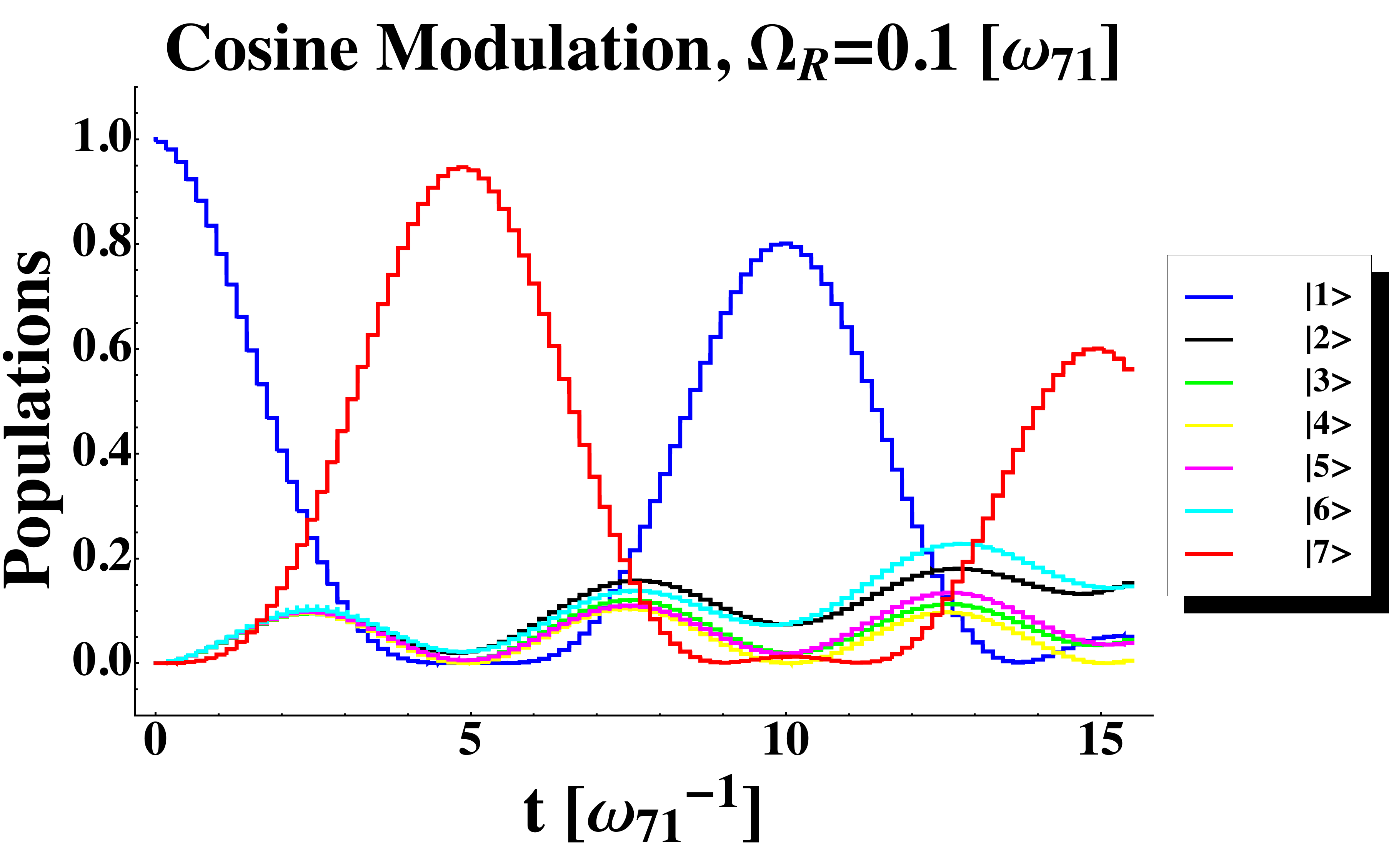}}
\caption{Population dynamics in the seven-level $\Lambda$ system induced by a cosine-modulated pulse train described by Eq.(\ref{OFC}), ($\phi=\pi/2$). The values of the system parameters are  $\omega_{32}$ = 410.7 THz, $\omega_{21}$ = 340.7 THz, $\omega_{31}$ =70 THz and $\Delta \omega = 0.1\omega_{31}$ \cite{Sh08}. The carrier frequency $\omega_0$ = $\omega_{32}$, the spectral modulation parameter $T_0$ = 1/$\omega_{21}$, the modulation amplitude $A=4$ and the peak Rabi frequency $\Omega_R$=7 THz. A single pulse duration $\tau$ = 3 fs, pulse train period is T = 19.2 ns. 
}\label{7lvl_cos}
\end{figure}

For the sine modulation, Fig.(\ref{dress_sinsptr}(a)), dynamics occurs within a single dressed state $|I>$ (blue) performing the adiabatic passage from the initial bare state $|1>$ (dashed blue) to the final bare state $|3>$ (dashed black). 
In contrast, the cosine modulation of the pulse train does not induce destructive interference of the spectral components actively involved into two-photon Raman transitions, resulting in no quasi-dark excited state generated. This phenomenon caused qualitatively different response of the seven-level system: population dynamics is shown in Fig.(\ref{7lvl_cos}) with a significant population accumulation in the transitional, excited state manifold. In the dressed state analysis Fig.(\ref{dress_sinsptr}(b)), a generalized adiabatic passage is also performed by a single dressed state $|I>$ (blue). However, the energy of the dressed state $|I>$ (blue) initially goes along with the bare state $|1>$ (dashed blue), then with the excited bare state $|2>$ (dashed green), and at the end of pulse duration coincides with the final bare state $|3>$ (dashed black). {\em The involvement of all three bare states in the dressed state dynamics is the reason for the excited states being populated under cosine modulation} in the exact solution in Fig.(\ref{7lvl_cos}). 
\begin{figure}
	\centerline{
		\includegraphics[width=9cm]{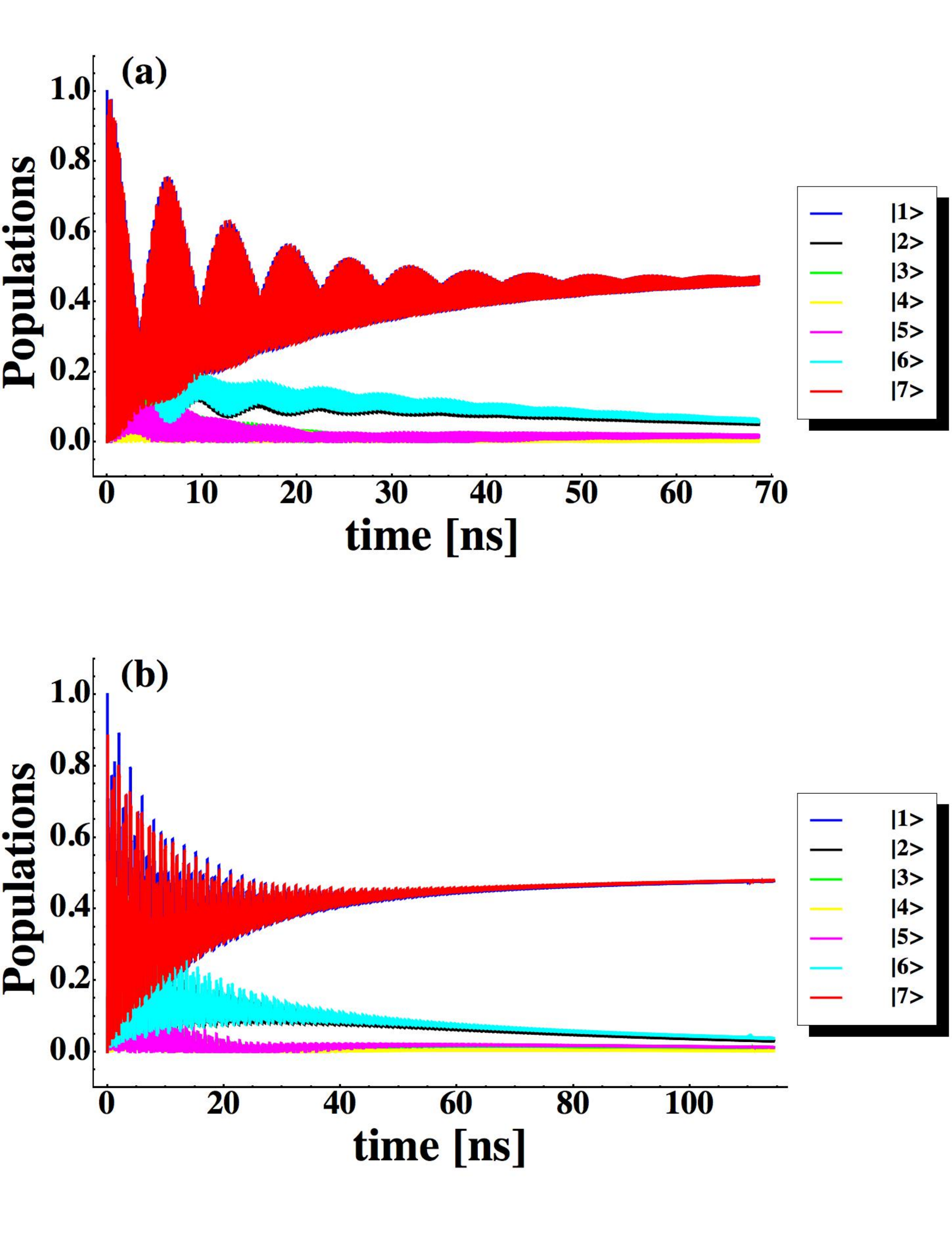}}
	\caption{Population dynamics in the seven-level $\Lambda$ system induced by a sine-modulated (a) and cos-modulated (b)  pulse train in the presence of decoherence. The values of the system parameters are  $\omega_{32}$ = 410.7 THz, $\omega_{21}$ = 340.7 THz, $\omega_{31}$ =70 THz and $\Delta \omega = 0.1\omega_{31}$ \cite{Sh08}. The carrier frequency $\omega_L$ = $\omega_{32}$, the spectral modulation parameter $T_0$ = 1/$\omega_{21}$, the modulation amplitude $A=4$ and the peak Rabi frequency $\Omega_R$=70 THz. A single pulse duration $\tau$ = 3 fs, pulse train period is T = 19.2 ns. All the spontaneous decay rate are the same which is $ \gamma=10^8\ s^{-1} $ and all the collision rate are the same which is $ \Gamma = 10^4\ s^{-1}$.
	}\label{sin_deco}
\end{figure}

Fig.(\ref{sin_deco}) shows population dynamics in the seven-levels system induced by the (a) sin- and (b) cos-modulated pulse train in the presence of decoherence. The spontaneous decay rate is $ \gamma=10^8\ s^{-1} $ and the collision rate is $ \Gamma = 10^4\ s^{-1}$ relevant for the ultracold temperatures regime \cite{Cu05}. Both figures demonstrate an asymptotically equal population between the initial Feshbach and the final ultracold states regardless to the parity of the chirp. This result is in contrast to the case of three-level $\Lambda$ system interaction with the pulse train under the same conditions \cite{Ma13}. The difference is due to the increased number of the excited states that effectively accelerate the decay rate, such that, even for an incremental population transfer to the excited state manifold, population spontaneously decays back to the initial state fast enough preventing its efficient accumulation in the final state.
In summary, a new method for controlled formation of molecules in the ultracold state is proposed implementing a single, sinusoidally modulated optical frequency comb. The induced dynamics leads to effective accumulation of molecules in the ultracold state on the time scale faster than the spontaneous decay time, within a few nanoseconds range. The parity of the spectral chirp shown to play a key role in the adiabatic passage, confirming previous investigations \cite{Ma13,Go02}.

\section{Funding Information}
This research is supported by the National Science Foundation
under Grant No. PHY-1205454.






\end{document}